# Radial deformation measurements of isolated pairs of single-walled carbon nanotubes

Z. R. Abrams and Y. Hanein*

*School of Electrical Engineering, Department of Physical Electronics, The Iby and Aladar Fleischman Faculty of Engineering, Tel-Aviv University, Tel-Aviv, Israel, 69978*

\* *Corresponding author. Email: [hanein@eng.tau.ac.il](hanein@eng.tau.ac.il). Tel: 972-3-6407698, Fax: 972-3-6423508*

Abstract:

Interactions between atoms of bound single-walled carbon nanotubes (SWNTs) are known to cause measurable distortion to the tube's original circular cross-section frame. High-resolution transmission electron microscope (TEM) investigation was used here to directly image and verify these radial deformations. The data presented here provide, for the first time, direct measurements of the deformations due to the interactions between isolated pairs of nanotubes. The deformation data is compared to previously reported experimental and simulation results.

PACS number: 61.48.+c, 61.16.Bg, 62.20.Fe

# I. INTRODUCTION

Carbon nanotubes (CNTs) are a stable allotrope of carbon that, due to their unique atomic makeup, have extraordinary mechanical, electrical, and chemical properties[1, 2]. CNTs can be ideally described as rolled-up graphene sheets, with a single-walled nanotube (SWNT) comprising of an individual graphene sheet, and a multi-walled nanotube (MWNT), consisting of nested layers of rolled-up graphene. The nanoscaled dimensions of these tubes imply that the van der Waals interactions between the CNT's atoms and its environs are of crucial importance[3, 4]. These interactions include both surface-tube and tube-tube interactions, and are especially relevant in the case of bundles of CNTs[5, 6]. A direct outcome of these interactions is the deformation of the nanotubes.

The deformation of a cylindrical cross-sectioned nanotube causes its various physical properties to change, most notably, its electrical[7] and mechanical[8] properties. Precise knowledge of these changes, backed by experimental results, is crucial in understanding the precise physics of CNT systems, whether in the scope of band-gap engineering of CNTs[9], or in structural analysis of CNT ropes[6].

While the radial deformations of MWNTs have been previously shown in Refs. 4 and 10, and numerous simulations of these results have been implemented[5, 6, 7], few direct studies of SWNT deformations exist[11]. Moreover, many of the methods applied in the past have relied on indirect means of studying these deformations. The majority of these studies make use of atomic force microscopy[3, 12, 13]. The disadvantage of this method is twice-fold: the first is that the lateral resolution of an atomic force microscope (AFM) is limited by the shape of the tip and the need to calculate the de-convolution between the tip end and the topography being measured, and the second, being that regardless of the working mode of the AFM, the technique is relatively invasive, affecting the CNTs while simultaneously measuring them. This latter point nevertheless has its advantages in terms of dynamic-response measurements of radial deformations in CNTs.

TEM analysis of CNT deformations, on the other hand, has the advantage of being both high-resolving, with the resolution limit dependant on the focus of the image, and non-invasive, with the CNTs being imaged in the static position in which they are grown. Nevertheless, nearly all TEM verifications of radial deformations in CNTs have been based upon MWNT studies[4], which complicate the matter from an atomic-modeling point of view. The only TEM verification of SWNT deformation has been in large bundles of multiple tubes, which form a hexagonally-packed array of deformed tubes[11].

Here, we utilize a unique form of growth and imaging technique whereupon we grow the CNTs *directly* onto a TEM nitride mesh having large, 40μm×40μm, or 2μm diameter, holes[15] (see Fig. 1), without the need for the preliminary growth of CNTs on a surface, followed by their deposition on a carbon grid. The scheme used here avoids the harsh sonication or other removal processes needed to extract the CNTs from the surface to which they are grown, a process that can add a multitude of defects to the CNTs, including the possible twisting and flattening of wide tubes[9, 14]. Moreover, carbon grids provide limited imaging contrast when compared to the carbon-based walls of a CNT. Our direct growth process is advantageous in many ways, including the ability to measure in-plane CNT growth, with maximal contrast[16], as well as simplifying the process of locating suspended CNTs.

The results of this work provide a first direct examination of the statics of pairs of nanotube interactions. Each image is a case study in itself, and the various parameters contained in each of these were taken into account, as well as providing the basis for a more holistic approach to CNT deformations.

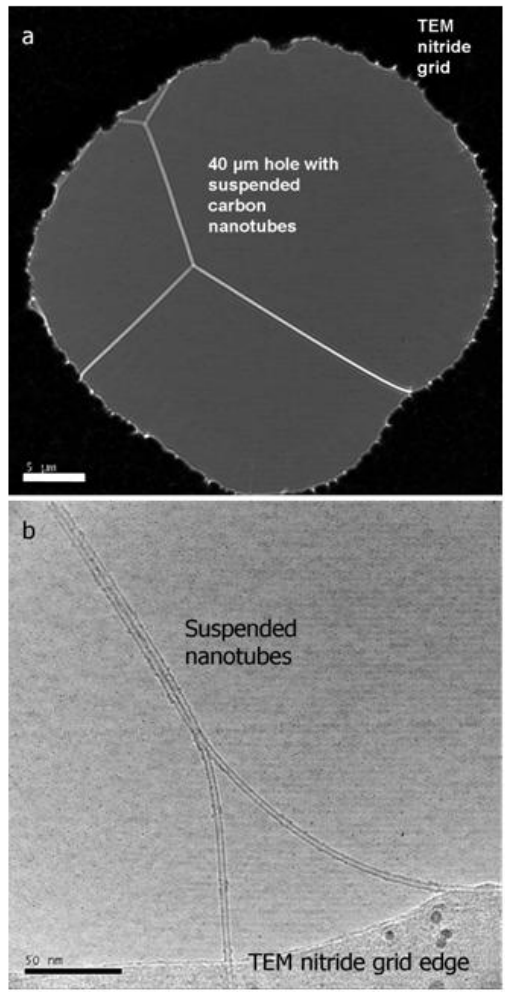

FIG. 1. (a) TEM image of suspended nanotubes grown over one of the large holes in the TEM nitride grid used (see text) [scale bar: 5μm]. (b) High-resolution TEM image of a pair of suspended nanotubes growing off the nitride grid [scale bar: 50nm].

## II. EXPERIMENTAL

CNTs were grown using the CVD method on TEM grids consisting of holey meshes in silicon nitride thin films (200 nm thick) held by silicon holders (DuraSiN mesh). The two types of meshes used in this study had arrays of either 40μm square holes (as in Fig. 1(a)), or 2μm diameter circles, in the nitride films. These specific films are robust enough to withstand the 900°C required in the CVD growth. Furthermore, the

CNTs grow suspended over the holes, as is described elsewhere by us[15], allowing the maximal contrast possible with a TEM, as opposed to CNTs deposited on carbon films.

The grids were prepared by dipping them in a suspension of iron nanoparticle catalysts. Briefly, granules of iron salt ($Fe_2(NO)_3$) were dispersed in 20cc of isopropyl alcohol, followed by sonication (40 min), and centrifuge (10 min). A drop of the suspension was cast directly onto the TEM grid, which was then promptly dipped in hexane, and dried with nitrogen. The sample was then placed in an automated CVD system which consists of a tube-furnace, gas-flow meters and a control computer. The feed gas for the growth was ethylene, at 900°C for 9 minutes, and the system was kept in a constant flow of hydrogen gas throughout the growth.

TEM analysis was done using a Philips Tecnai F20 TEM, at 200kV, and all measurements of the images was facilitated by a commercial imaging program (Motic Images Plus). All measurements were based on averaging at least two cross-sectional lines; the error in these measurements is based upon the pixelization created by the graphics file format, and the limiting resolution of the TEM. The standard deviation from the average diameter was no more than 5% per tube. Almost all adjoining pairs of CNTs were seen to deform to an extent, however, only those cases where the *exact* diameter of the SWNT was measurable were used for analysis. Data from images consisting of nanoubes adhering to the edge of the nitride grid (tube-surface interactions) are inconclusive, as the nitride is semi-covered in carbon due to the CVD growth, and thereby creates gradients in the radial deformation distribution along the surface-tube contact area.

## III. RESULTS AND MEASUREMENTS

Figure 2 displays two examples of CNT adherence at a junction point, revealing conspicuous variations in tube diameter. Clearly, the deformations viewed using the TEM are dependant on the viewing angle. Two distinct types of deformation presented in Fig. 2 are those viewable as top-down images, whereupon the diameters of the tubes elongate due to the "squishing" (van der Waals) forces; and side-viewed images (in

Fig. 2(a) only), where the deformed tube will contract due to the same forces. Other types of images (not shown) are a superposition of these two types of views; however, the amount of deformation viewable in such a combination will not give the maximal deformation, and are thus less preferred. It must also be noted that as the TEM images are two-dimensional renderings, the term 'above' and 'below' are interchangeable when describing the top-down images.

Figure 2(a) exemplifies *both* types of deformation images. The relatively straight nanotube undergoes both types of deformation along the interaction surface. The radial deformation is distinctly evident in both forms. Fig. 2(b) presents another case of distinct elongation of the nanotube's diameter. Furthermore, in this image, the diameters of both CNTs are seen to widen at the interaction surface. Nevertheless, special care must be taken when analyzing such an image, as the upper CNT in this image is a double-walled CNT (DWNT), complicating the analysis. It should be noted that the original diameters of the tubes are all measured far from the junction point. Additionally, the buckling of the tubes, as opposed to their gradual curving has previously been analyzed by us[15]. The buckled tube displayed here deforms more than a tube of similar dimensions, as mentioned in Ref. 3, and can therefore be seen as an extreme case of deformation.

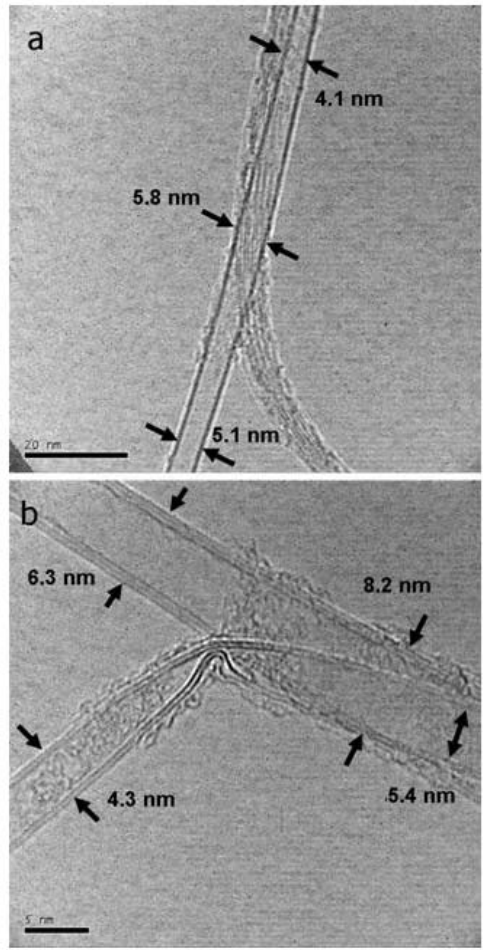

FIG. 2. TEM images of nanotube junctions showing their radial deformation: (a) Top-down view of a single SWNT deformed in both radial directions due to interactions with both the top of the underlying CNT bundle, for which the diameter elongates, and with the side of the bundle, for which it contracts [scale bar: 20nm]. (b) A CNT buckling junction. The bottom SWNT undergoes both a buckling, and a radial deformation [scale bar: 5nm]. The top CNT is a Double-Walled CNT, and also widens [*]. All measurements are averages of cross-sections.

Figure 3 illustrates more complex examples of diameter deformations. In Fig. 3(a), a side-view of a SWNT adjoined to a bundle of other SWNTs also shows deformation, however, while the contraction of the tube is visible, the amount of distortion may be attenuated by the other SWNT running below (or, above) the junction point. Furthermore, as the bundle to which the SWNT here is adhering is relatively large, the surface area between the deformed SWNT's wall and the bundle's walls is larger than a simple tube-tube interaction, and more comparable to a nanotube adhering to a surface, as will be discussed later in the text. Further corroboration of this deformation is achievable when compared to the method of Ref. 4, as the distinctly darker fringe near the buckle/adhesion point alludes to the flattening of the CNT wall at the binding surface, creating a thicker column of carbon atoms for the electrons to deflect off. In Fig. 3(b), the SWNT can be seen to elongate a little; however, the deformation is not as much as in similar examples (i.e. those of Fig. 2). In addition, the second tube is both smaller and is not completely parallel to the first tube, minimizing the effective surface area of the tube-tube interaction.

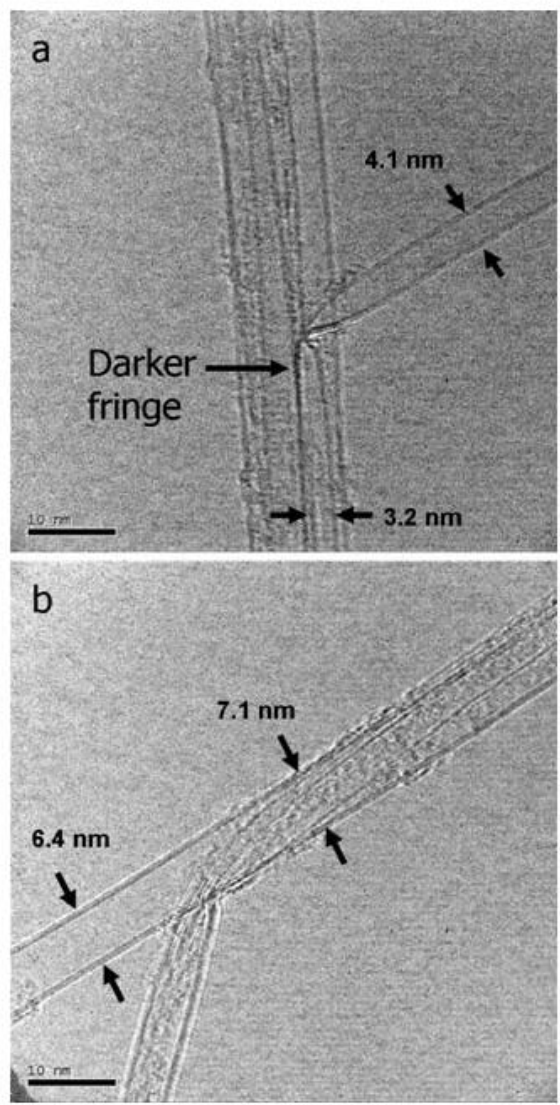

FIG. 3. TEM images of more complex systems of nanotubes. (a) A SWNT buckling to a bundle of other SWNTs. The side-view deformation is visible, including the flattening of the nanotube wall [scale bar: 10nm]. (b) A top-down view of a SWNT with a smaller CNT running beneath it. Here, the surface area between the tubes changes along the length of the junction [scale bar: 10nm].

The images presented here demonstrate the complexity of each individual nanotube system, as an isolated case study. Table I summarizes some of the types of radial deformation information gleaned from the TEM images in Figs. 1, 2 and 3, including some of the parameters needed to be taken into consideration in their analysis. While not displayed in the image itself, the pair of CNTs adhering to each other in Fig. 1(b) also underwent radial deformation after the y-junction. The two tubes are both DWNTs of similar diameter (3.4, 3.6 nm, information from close-up image not shown) and the two tubes remain completely parallel after binding, with no further twisting. The images are labeled in Table I according to the order presented in the text. Images similar to the 'Sample image 2' (i.e. Fig. 2(a), see Table I) are the most prevalent form, with the top-down images taken as the more reliable (when compared to side-viewed ones). In addition to the data taken from the images presented here, Table I also includes three more data points measured (images not shown). These additional data points are discussed later in the text.

Table I: Summary of Deformation Data and Parameters

| Sample image | Free tube diameter [nm] | Deformed (bound) tube diameter [nm] | % deformation | Side/Top view | Diameter of adjoining tube/tubes [nm] | Other parameters |
|---|---|---|---|---|---|---|
| 1 [1b] | 3.6, 3.5 | 4, 3 | -16%, -15% | Side | - | Two similar DWNTs binding |
| 2 [2a] | 5.1 | 5.8 | 15% | Top | ~6.5 | Attached to bundle |
| 3 [2a] | 5.1 | 4.1 | -20% | Side | ~6.5 | Attached to bundle |
| 4 [2b] | 4.3 | 5.4 | 27% | Top | 6.3 (#6) | DWNT |
| 5 [2b] | 6.3 | 8.2 | 30% | Top | 4.7 (#5) | Buckled |
| 6 [3a] | 4.1 | 3.2 | -23% | Side | - | Buckled, attached to thick bundle |
| 7 [3b] | 6.4 | 7.1 | 10% | Top | ~4.6 | Adjoining tube deviates from tube axis |
| 8 | 2 | 2.1 | 4% | Top | 4.5 | Slight buckle |
| 9 | 3.7 | 4 | 7% | Top | ~4.5 | Attached to bundle of two tubes |
| 10 | 5.6 | 6.7 | 20% | Top | ~6 | Attached to bundle |

## IV. DISCUSSION AND ANALYSIS

The results presented above illustrate the effectiveness of our TEM technique in studying the radial deformations of CNTs. The method benefits from several major advantages: The first lies in the geometry of the TEM grids, which confines the CNTs to the same growth and imaging plane. The second is in the verification of CNTs reverting back to their original diameter in the absence of external force; specifically, images of CNTs binding in short segments, and then splitting off, show that the free ends of the SWNTs have the same diameter on either end of the junction. This is perhaps the foremost advantage of the TEM grid system used here, as both the original and distorted diameters of the suspended nanotubes can be measured simultaneously, without surface-tube interactions, as in AFM studies. Finally, as the CNT systems analyzed in this study were all produced during the CVD growth, they can consequently be seen as a local minimal energy configuration of the CNTs, with the radial deformations imaged in their static position. Thus, we can utilize our results for exploring the underlying physical properties associated with the deformations.

The first issue quantifiable with our data is in measuring the extent of these deformations as a function of the nanotube's diameter. To quantify the extent of these deformations, we plotted the relevant data presented in Table I in Fig. 4. Figure 4 displays two sets of TEM data. The first set (set 1) corresponds to data points from top-down TEM images only, in which a single SWNT is seen to deform (Sample images:

3, 8-10, from Table I), and the second set corresponds to data taken specifically from Fig. 2(b). The data from set 1 portrays a monotonic rise in deformation of the SWNTs as a function of increasing diameters. The results of our TEM measurements in set 1 are consistent with the previously measured and simulated results by Hertel *et al.*[3], also plotted in Fig. 4, in which the calculated deforming effects were due to tube-surface interactions. The tube-surface simulated data from Ref. 3 acts as an upper bound to the measured TEM data presented here in set 1, since the interactions between a tube and the surface is more than that of two individual, bound tubes, causing less deformity to the tubular cross-section. Also apparent is the fact that the amount of deformation is attenuated if the CNT in question is a MWNT (shown in Fig. 4 by the open square). The slope of the curve of set 1 in Fig. 4 should be factored relative to the size of the adjoining tube. Specifically, it should be noted that the data points are related to CNTs interacting with other CNTs of different diameters (4.5, 4.5, 6.5, and 6 nm, from left to right respectively).

The data in set 2 of Fig. 4 (open circles) represents a more complex set of data, which nevertheless substantiate the consistency between the results presented in this study and those of the simulated tube-surface data. The rightmost data point is that of the larger (top) DWNT in Fig. 2(b); alongside it (isolated open square) is the simulated deformation of a 5.4nm (40, 40) tube[3]. As simulated, the deformation is lowered in respect to the simple case of SWNT-surface interaction (closed squares) since the double-walled tube is harder to deform. The left point of set 2 (leftmost open circle) corresponds to the buckled tube of Fig. 2(b). The data point for this tube-tube derived deformation is very close to the upper bound of the tube-surface simulated result. Further analysis of Fig. 2(b) suggests that the smaller (bottom) tube (leftmost open circle) in the image perceives an almost flattened layer of graphite due to the reciprocal flattening of the larger (upper) tube (rightmost open circle). The strong deformation of this tube may also be influenced by the fact that buckled tubes are more easily deformed[3].

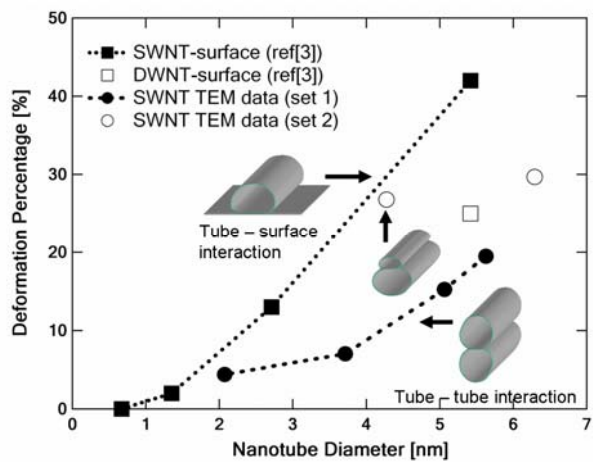

FIG. 4. Tube deformations due to tube-surface and tube-tube interactions, plotted versus deformed nanotube diameter. The data from the measured TEM results is compared to that of Hertel et al., *Physical Review B*, **58**, 13870 (1998). The Squares are simulated, tube-surface, results, whereas the Circles are experimentally measured ones. The open square refers to a double-walled nanotube, whereas the closed squares refer to single walled nanotubes. The TEM data from set 1 is taken from Table I (Sample images 3, 8-10), and set 2 is that of Fig. 2(b). Included, is a rendering of the types of deformations listed here. The upper-left is that of a flattened tube on the surface, as depicted by the upper bound curve, which is that of maximal deformation. The bottom-right is of the typical type presented in this study. The central presents a smaller tube adhering to a larger one, and is deformed almost as if it were bound to a wide surface, as in Fig. 2(b), this explains why the point lies so close to the simulated curve.

Overall, the data presented above emphasizes the various parameters involved (viz. deformed tube diameter, interacting tube diameter, number of walls) complicating the simple molecular dynamics simulations generally implemented in describing the mechanics of either CNT ropes[17], or in measurements of the binding energy between nanotubes[18]; specifically, the deformation of the nanotubes along one another, in conjunction with the varying radial distribution of CNTs comprising these systems. An additional consequence of the presented data is the need to consider the details of the anisotropic deformation when attempting to analyze the exact properties of deformed tubes[19].

A second quantifiable issue is the binding energy between nanotubes. As has been previously demonstrated[3, 11, 20], the deformation of the SWNTs has a major effect on the value of the binding energy between nanotubes. Using the data acquired here regarding the deformation of tubes, a more precise measurement of the binding energy between nanotubes can be evaluated, similar to the analysis done in

Ref. 3. A more detailed account of binding energy measurements is beyond the scope of this publication and will be discussed by us in the future.

## V. CONCLUSION

The data presented here reveals, for the first time, the distinct radial deformation of individual SWNTs. The analysis can be done visually, due to the use of the high-resolution TEM imaging technique used[16]. This forgoes the need for external force[3], diffraction analysis[4], or moiré pattern detection[21]. The data shows the various parameters needed to measure the effect of each nanotube's deformation, with an emphasis on the dependence of the deformed nanotube, and its diameter. The data is also shown to be consistent with existing simulation results[3, 18] of nanotube deformation due to tube-surface interactions.

Whether isolated, or in bundles, SWNTs are seen to dramatically alter their dimensions in response to their binding to both the surface, and other tubes. Taking a CNT's deformation into account is critical if they are to be used in electrical, mechanical, or nano-electro-mechanical systems (NEMS).

## ACKNOWLEDGEMENTS

The authors thank Dr. Y. Lereah for his assistance with the TEM, and Dr. R. Lifshitz for his helpful input. This research was supported by an ISF grant.

[1] Carbon Nanotubes Synthesis, Structures, and Applications, edited by M. S. Dresselhaus, G. Dresselhaus, and Ph. Avouris (Springer, Berlin, 2001).

[2] H. Dai, *Surface Science* , **500**, 218 (2002).

[3] T. Hertel, R. E. Walker, P. Avouris, *Physical Review B*, **58**, 13870 (1998).

[4] R. S. Ruoff, J. Tersoff, D. C. Lorents, S. Subramoney, and B. Chan, *Nature*, **364**, 514 (1993).

[5] M. J. López, A. Rubio, J. A. Alonso, L.-C. Qin, and S. Iijima, *Physical Review Letters,* **86**, 3056 (2001).

[6] D. Qian, W. K. Liu, R. S. Ruoff, *Composites Science and Technology*, **63**, 1561 (2003).

[7] A. Rochefort, P. Avouris, F. Lesage, and D. R. Salahub, *Physical Review B*, **60**, 13824 (1999).

[8] A. Pantano, D. M. Parks, and M. C. Boyce, *Journal of the Mechanics and Physics of Solids*, **52**, 789 (2004).

[9] O. Gulseren, T. Yildirim, S. Ciraci, and C. Kılıç¸ *Physical Review B*, **65**, 155410 (2002)

[10]V. Lordi and N. Yaoa, *Journal of Chemical Physics*, **109**, 2509 (1999).

[11] J. Tang, L. -C. Qin, T. Sasaki, M. Yudasaka, A. Matsushita, and S. Iijima, *Physical Review Letters*, **85**, 1887 (2000).

[12] M. –F. Yu, T. Kowalewski, and R. S. Ruoff, *Physical Review Letters*, **85**, 1456 (2000).

[13] R. S. Ruoff, D. Qian, and W. K. Liu, *C. R. Physique*, **4**, 993 (2003).

[14] N. G. Chopra, L. X. Bennet, V. H. Crespi, M. L. Cohen, S. G. Louie, and A. Zettl, *Nature*, **377**, 135 (1995).

[15] Z. R. Abrams and Y. Hanein, *cond-mat/0605696* (2006).

[16] Z. R. Abrams, Y. Lereah, and Y. Hanein, *In preparation* (2006).

[17] Z. Liu, and L.–C. Qin, *Carbon*, **43**, 2146 (2005).

[18] B. Chen, M. Gao, J. M. Zuo, S. Qu, B. Liu, and Y. Huang, *Applied Physics Letters*, **83**, 3570 (2003).

[19] S. Komura, K. Tamura, and T. Kato, *European Physical Journal E*, **13**, 73 (2004).

[20] T. Tang, A. Jagota, and C. –Y. Hui, *Journal of Applied Physics*, **97**, 074304 (2005).

[21] H. Shang, H. Xie, H. Zhub, F. Dai, D. Wu, W. Wang, and Y. Fang, *Journal of Materials Processing Technology*, **170**¸ 108 (2005)

[*] An interesting, unrelated, feature apparent in these images is the appearance of collateral material adjacent to the CNTs near the junction points. This is easily discernible in Figs. 2(a) and (b), and, to a lesser extent, in Figs. 3(a) and (b). While the common response to the puzzle as to what this stray material is usually is carbon, there is nothing to deny the existence of other materials adhering to the CNTs. EDX of these junctions does not provide any additional information due to the lack of response to low atomic-numbered materials (EDX analysis showed carbon in large amounts, and oxygen in smaller amounts, perhaps implying the adherence of water molecules). The strain present in the bent or buckled CNT walls,

caused by both the adherence to another tube, and the radial deformation thereby induced, may further hybridize the σ and π bonds, which perhaps leads to a stronger affinity for external adhesion of particles.